# The Dynamics of Drop Breakup in Breaking Waves

W.H.R. Chan (Center for Turbulence Research, Stanford University, USA;
Present Address: University of Colorado Boulder, USA)

**ABSTRACT**

Breaking surface waves generate drops of a broad range of sizes that have a significant influence on regional and global climates, as well as the identification of ship movements. Characterizing these phenomena requires a fundamental understanding of the underlying mechanisms behind drop production. The interscale nature of these mechanisms also influences the development of models that enable cost-effective computation of large-scale waves. Interscale locality implies the universality of small scales and the suitability of generic subgrid-scale (SGS) models, while interscale nonlocality points to the potential dependence of the small scales on larger-scale geometry configurations and the corresponding need for tailored SGS models instead. A recently developed analysis toolkit combining theoretical population balance models, multiphase numerical simulations, and structure-tracking algorithms is used to probe the nature of drop production and its corresponding interscale mass-transfer characteristics above the surface of breaking waves. The results from the application of this toolkit suggest that while drop breakup is a somewhat scale-nonlocal process, its interscale transfer signature suggests that it is likely capillary-dominated and thus sensitive not to the specific nature of large-scale wave breaking, but rather to the specific geometry of the parent drops.

**INTRODUCTION**

Breaking waves on the surfaces of oceans generate drops of a broad range of scales (de Leeuw *et al.*, 2011; Veron, 2015; Wang *et al.*, 2016; Erinin *et al.*, 2019; Deike, 2022; and references therein). These drops have a significant influence on weather- and climate-relevant ocean–atmosphere interactions, including the enhancement of near-ocean-surface mass, momentum, and energy transfer (Andreas, 1992; Andreas *et al.*, 1995; Veron, 2015; Deike, 2022), the distribution and activation of organic material (Cunliffe *et al.*, 2013; Deike, 2022), as well as the production of nucleation sites for cloud formation (Tao *et al.*, 2012; Veron, 2015; Fan *et al.*, 2016). The latter is especially relevant in the naval context since ships have been observed to generate trails of thunderclouds in their wake (Chang, 2017). Near-surface air–sea fluxes also modify the atmospheric and oceanic boundary layers with significant impact on ship operation and performance (Andreas and Emanuel, 2001). Understanding the generation of these ship trails and boundary-layer modifications, and more generally the impact of drops and their successors on our weather, climate, and environment, require a thorough characterization of the dynamics of drop formation and evolution.

The drop size distribution has been directly characterized by a number of laboratory and field experiments (Wu, 1979; Koga, 1981; Wu *et al.*, 1984; de Leeuw, 1986; Smith *et al.*, 1993; Anguelova *et al.*, 1999; de Leeuw *et al.*, 2000; Fairall *et al.*, 2009; Veron *et al.*, 2012; Erinin *et al.*, 2019, 2022; Mehta *et al.*, 2019), as well as recent numerical simulations (Wang *et al.*, 2016; Mostert *et al.*, 2022). These results offer a glimpse into the physical mechanisms behind drop production. We have recently developed an analysis toolkit that probes the causes behind the dynamic evolution of the drop size distribution through a combination of theoretical models based on population balance analysis, numerical simulations using an interface-capturing method, and structure-tracking algorithms to identify and track individual bubbles/drops and their associated breakup events (Chan, 2020; Chan *et al.*, 2021abc). This toolkit was successfully used to probe the interscale nature of turbulent bubble breakup in breaking waves, but is not restricted to turbulent bubble fragmentation and may be used to analyze a variety of multiphase flows in different contexts. In this work, the interscale nature of drop breakup above the surface of breaking waves is investigated using the same analysis toolkit.

The objectives of this study are to utilize a combination of theoretical analysis and numerical simulations to probe the fundamental mechanisms behind drop production in breaking waves by analyzing the transfer of liquid mass within drops between different drop sizes through the occurrence of breakup events. The theoretical nature of the interscale mass flux between drop sizes is specifically examined. The interscale locality of this flux determines the universality of the underlying drop breakup mechanism and informs strategies for SGS model development to enable cost-effective computation of large-scale waves. A scale-local flux implies small-scale universality and the feasible application of universal SGS models, while a scale-nonlocal flux suggests that models may have to be tailored to the specific parameters of the flow since breakup is then likely to be geometry dependent. The structure of this paper is as follows: the employed methodology is introduced, results obtained from application of the analysis toolkit are discussed, and a summary and outlook for future work are offered in the conclusions.

## METHODOLOGY

The methodology employed in this work is discussed in detail by Chan (2020) and Chan *et al.* (2021abc). Key aspects of the methodology are recapitulated here.

### Analysis toolkit workflow

Our analysis toolkit may be applied as follows: first, an ensemble of numerical simulations of the flow of interest is performed to resolve bubble and/or drop breakup and/or coalescence processes and to obtain sufficiently converged bubble and/or drop statistics in the range of scales of interest. Second, structure-tracking algorithms are applied to identify and track all resolved bubbles and/or drops in the system. Bubble and/or drop lineages are constructed and used to identify breakup and/or coalescence events, which are then used to evaluate the interscale bubble-mass and/or droplet-mass flux. Third, the properties of the measured flux are extracted and used to validate existing theories or construct possible mechanisms for bubble and/or drop formation. The cornerstone of this analysis toolkit is the interscale mass flux, which will be introduced next in the context of drop breakup.

### The interscale droplet-mass flux

The interscale droplet-mass flux due to binary breakup events, $W_b(D)$, is defined as the rate of transfer of liquid mass present in the form of drops that is lost by all drops of sizes larger than a particular cutoff value $D$ and gained by all drops of sizes smaller than $D$. By the principle of mass conservation, these two quantities are equivalent. The breakup flux $W_b(D;t)$ at a particular time $t$ may be expressed as

$$W_b(D;t) = \int_0^D dD_c \, D_c^3 \times \\ \times \int_D^\infty dD_p \, q_b(D_c;t|D_p) g_b(D_p;t) f(D_p;t), \quad (1)$$

where $D_c$ and $D_p$ are integration dummy variables representing, respectively, the child and parent drop sizes, $q_b(D_c;t|D_p)$ is the probability distribution of sizes of child drops given a particular parent drop size $D_p$, $g_b(D_p;t)$ is the characteristic breakup rate of drops of size $D_p$, and $f(D_p;t)$ is the drop size distribution of the droplet population.

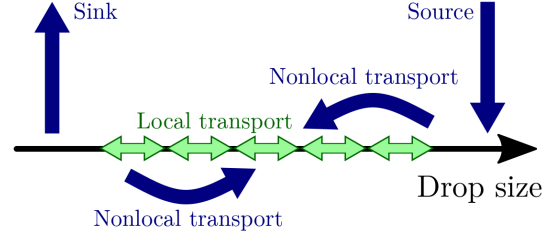

**Figure 1:** Schematic of potential contributions to the interscale droplet-mass flux as visualized in drop-size space.

The interscale flux $W_b(D;t)$ is an amalgamation of several possible contributions to the rate of change of the drop size distribution $f(D;t)$. Figure 1 illustrates the nature of these contributions. The evolution of $f(D;t)$ may be expressed in terms of the phenomenological population balance equation

$$\frac{\partial[f(D;t)D^3]}{\partial t} = \frac{\partial W_b(D;t)}{\partial D} = \\ = \frac{\partial W_{b,\text{local}}(D;t)}{\partial D} + T_{\text{source}}(D;t) - T_{\text{sink}}(D;t). \quad (2)$$

Here, we are neglecting the contributions of nonbreakup events, such as coalescence events, to the drop population. The effects of coalescence are deferred to future work. Scale-nonlocal transport carries droplet mass from one drop size to a distinct, nonneighboring drop size and may be expressed as a summation of discrete sources $T_{\text{source}}$ and sinks $T_{\text{sink}}$. The interscale flux due to scale-local transport, $W_{b,\text{local}}(D;t)$, may be expressed as

$$W_{b,\text{local}}(D;t) = v_D(D;t) f(D;t) D^3, \quad (3)$$

which is the scale-local flux of $fD^3$ in drop-size space ($D$-space) carried at a characteristic speed $v_D$. The speed $v_D$ may be interpreted as the time $t \sim 1/g_b$ taken for the droplet population to traverse a size range $\Delta D \sim D$. Observe that all the terms in Eq. (3) are dependent only on $D$ (and $t$). This is a hallmark of scale locality: the interscale flux across a certain cutoff size $D$ is only dependent on quantities evaluated at size $D$, and no other parameters. In particular, it does not depend on drop statistics at sizes much larger or smaller than $D$. If the actual interscale flux $W_b(D;t)$

were assessed to be highly local, then one may reasonably approximate

$$W_b(D; t) \approx W_{b,\text{local}}(D; t), \quad (4)$$

and the effects of drop breakup in drop-size space can be approximately described by a single quantity $v_D$.

To analyze interscale locality, the interscale flux $W_b(D; t)$ may be decomposed into its differential contributions from different parent drop sizes $D_p$, or its differential contributions to different child drop sizes $D_c$. The higher the degree of interscale locality in each of the following metrics, the more concentrated the differential contributions (integrands) are around the cutoff size $D$. The first metric, infrared locality, may be expressed in terms of the following integrand

$$I_p(D_p; t|D) = \\ = \int_0^D dD_c \, D_c^3 q_b(D_c; t|D_p) g_b(D_p; t) f(D_p; t) \quad (5)$$

such that

$$W_b(D; t) = \int_D^\infty dD_p \, I_p(D_p; t|D). \quad (6)$$

The interscale flux is highly infrared local if the integrand $I_p$ decays as $D_p$ increases away from $D$. The greater the decay rate, the higher the degree of infrared locality. Figure 2 illustrates this behavior in a highly infrared local process. In a completely infrared local process, $I_p$ is a Dirac delta function centered at $D$.

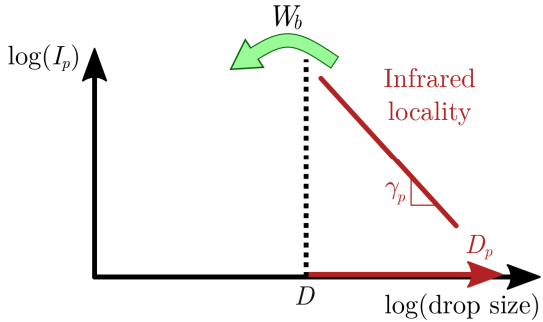

**Figure 2:** Schematic of infrared locality in drop-size space. $\gamma_p$ denotes the asymptotic decay exponent of $I_p$.

The second metric, ultraviolet locality, may be expressed in terms of the following integrand

$$I_c(D_c; t|D) = \\ = \int_D^\infty dD_p \, D_c^3 q_b(D_c; t|D_p) g_b(D_p; t) f(D_p; t) \quad (7)$$

such that

$$W_b(D; t) = \int_0^D dD_c \, I_c(D_c; t|D). \quad (8)$$

The interscale flux is highly ultraviolet local if the integrand $I_c$ decays as $D_c$ decreases away from $D$. The greater the decay rate, the higher the degree of ultraviolet locality. Figure 3 illustrates this behavior in a highly ultraviolet local process. In a completely ultraviolet local process, $I_c$ is a Dirac delta function centered at $D$. The analyses of infrared and ultraviolet locality provide two distinct, but complementary, ways of decomposing the interscale flux.

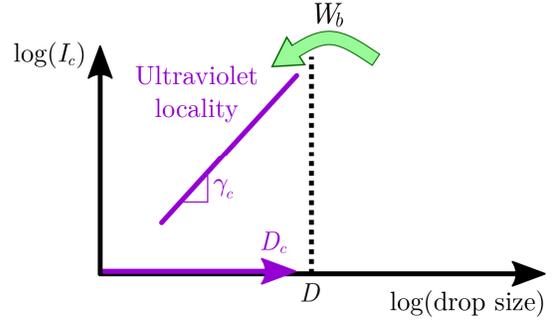

**Figure 3:** Schematic of ultraviolet locality in drop-size space. $\gamma_c$ denotes the asymptotic decay exponent of $I_c$.

Analytical scaling laws for $I_c$ and $I_p$ may be derived if the corresponding scaling laws for the constituent quantities $q_b$, $g_b$, and $f$ are known. For example, classical turbulence scaling laws for these constituent quantities were used to derive the desired scaling laws for turbulent bubble breakup. This is currently not as feasible for the case of drop breakup as the dominant underlying breakup mechanism remains a subject of active research. In this work, we focus on the measurement of $I_c$ and $I_p$ from numerical simulations and their relation to the observed scaling laws for $q_b$, $g_b$, and $f$, and defer a more detailed discussion of theoretical scaling laws to future work.

**Breaking-wave simulation ensemble setup**

Similar to the work described by Chan *et al.* (2018, 2019, 2021c), an ensemble of numerical simulations of breaking waves was generated to obtain the desired statistics. Details of the simulation parameters may be found in the aforementioned references. In particular, the simulations were initialized with a third-order Stokes wave of steepness 0.55. The Reynolds and Weber numbers of the wave, based on the wavelength $L$ and deep-water phase velocity $u_L = \sqrt{gL/2\pi}$ of the fundamental mode, are respectively $\text{Re}_L = 1.8 \times 10^5$ and $\text{We}_L = 1.6 \times 10^3$, matching those of a 27-cm-long water wave at atmospheric conditions. Note that these are at least one to three orders of magnitude smaller than those relevant to full-scale ship flows. However, the results of sufficiently small drops are expected to be relevant to all flows of similar configurations with sufficiently large $Re_L$ and $We_L$ if the interscale dynamics are sufficiently local, by invoking the hypothesis of (spatially) local isotropy commonly used in turbulence analysis. In these parameters, $g$ is the magnitude of standard gravity. All subsequent length scales are nondimensionalized by $L$.

A large ensemble size is required for statistical convergence, since the wave-breaking process is statistically unsteady, and time averaging is only valid within a limited time window in which a single breakup mechanism is known to dominate. As will be discussed later, the relatively small number of drops per realization compared to bubbles also necessitates a larger ensemble size in this work. Here, 250 independent wave realizations are used. A snapshot of the air–water interface from a representative wave realization is depicted in Figure 4.

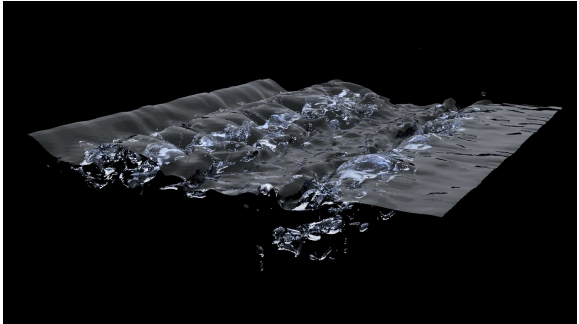

**Figure 4:** Snapshot of the air–water interface from a representative wave realization of the breaking-wave simulation ensemble discussed in the main text.

To enable this increased ensemble size compared to the previously referenced works, an algebraic volume-of-fluid method developed by Cascade Technologies, Inc. was used in contrast to the previously employed geometric volume-of-fluid method by the same developers (Kim *et al.*, 2021). Both are interface-capturing methods that directly resolve the air–water interface. The algebraic method exhibits improved efficiency and scalability as the need for geometric operations is eliminated, without sacrificing accuracy and predictivity as demonstrated by the verification and validation studies in the reference above. However, the slightly increased effective interface thickness in the algebraic method compared to the geometric method influences the identification of drops and the computation of the size distribution, as discussed later. In both methods, drops are counted only when their volume spans more than one cell, and the peaks in the distributions to follow (Figures 6 and 7) approximately mark the point where there are two cells across the drop diameter, so the distributions are reliable only to the right of this peak.

**Tracking of drop breakup events**

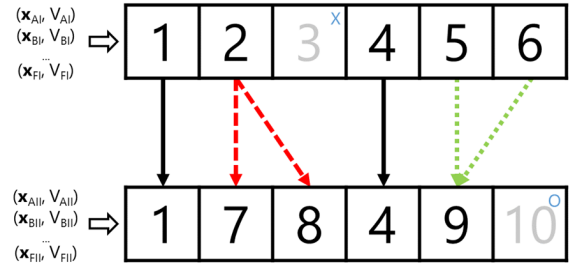

**Figure 5:** Schematic illustrating the drop tracking algorithm. Drop centroids and volumes from two consecutive snapshots are assembled and used to determine continuing drops (solid arrows), as well as breakup (dashed arrows), coalescence (dotted arrows), birth (O), and death (X) events.

Identification and tracking algorithms are applied to the previously described breaking-wave simulation ensemble to enable the accurate computation of drop volumes and centroids, which are then used to construct drop lineages, or "family trees", associated with the wave-breaking process. These lineages enable the detection of breakup and coalescence events. The algorithm requires a choice of the time interval between snapshots as illustrated in Figure 5. Two snapshot intervals are used in this work: $\Delta t_s = 6.0 \times 10^{-3}$ and $2\Delta t_s$. All times in this work, including these intervals, are nondimensionalized by $\sqrt{L/g}$.

**RESULTS**

In the plots that follow, an overbar denotes volume averaging over the computational domain of each realization and ensemble averaging over all relevant realizations, while the superscripts $T$ and $D$ denote averaging over time and size intervals to be specified, and a tilde denotes normalization by the maximum value reached in the plot of interest.

**Bubble size distribution**

We first present the bubble size distribution from the current simulation ensemble as a form of validation, as the power-law scaling for the size distribution is well established and our analysis toolkit has previously demonstrated the dominance of the breakup cascade mechanism underpinning this theoretical scaling law.

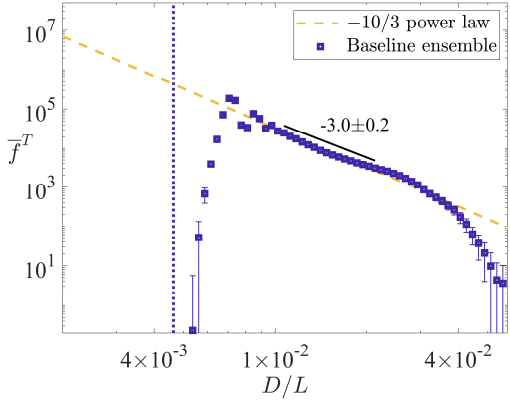

**Figure 6:** Bubble size distribution computed from the breaking-wave simulation ensemble, volume averaged over the domain of each realization, ensemble averaged over all relevant realizations, and time averaged over the interval [2.30, 3.14]. The vertical dotted line denotes the grid resolution, while the dashed line denotes the analytically derived and previously observed $D^{-10/3}$ power-law scaling. The uncertainties in the distribution and scaling exponent span two standard errors in each direction.

The computed distribution in Figure 6 agrees well with the analytically derived and previously observed $D^{-10/3}$ power-law scaling for turbulent bubble breakup in an intermediate range of scales (see Chan *et al.*, 2021c for relevant references). However, qualitative differences are observed between this distribution and our previously obtained distribution at large bubble sizes, likely due to the differences in the interfacial description in both numerical methods and their impact on the bubble identification algorithm. Investigations into these differences are under way.

**Drop size distribution**

We next present the drop size distribution from the current simulation ensemble in Figure 7. The computed distribution has some overlap with a previously observed $D^{-4.7}$ power-law scaling for large drops, which is midway between the $D^{-4.9}$ power-law scaling observed by Erinin *et al.* (2019) and the $D^{-4.5}$ power-law scaling computed by Wang *et al.* (2016). It also appears to be reasonably described by a gamma distribution fit (Villermaux, 2007). Observe that the number of drops is significantly lower than the number of bubbles at all sizes, necessitating a larger ensemble size for reasonable statistical convergence. Finally, note that the drop size distribution decays more rapidly than the bubble size distribution, and there are much fewer large drops than large bubbles. The smaller size range spanned by the drop population implies that the average parent drop size is much smaller than the average parent bubble size in breakup events.

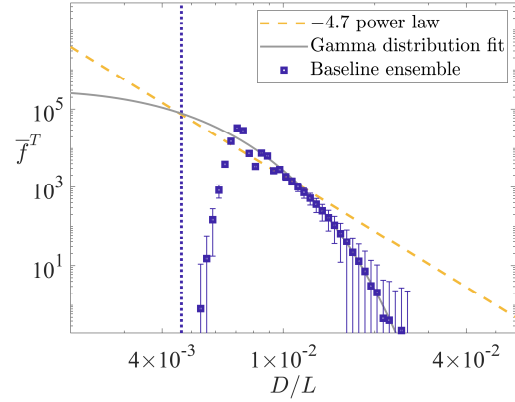

**Figure 7:** Drop size distribution computed from the breaking-wave simulation ensemble, volume averaged over the domain of each realization, ensemble averaged over all relevant realizations, and time averaged over the interval [2.30, 3.95]. The vertical dotted line denotes the grid resolution, while the dashed line denotes a $D^{-4.7}$ power-law scaling close to that observed in previous works. The uncertainties in the distribution span two standard errors in each direction.

It should be remarked that the drop size distribution turns out to have a higher grid sensitivity than the bubble size distribution for the resolution considered in this work. As discussed in an appendix, this grid sensitivity likely arises from delayed numerical breakup of liquid ligaments due to suppressed viscous drainage (lubrication flow), and suggests that the ensuing results only represent a subset of all physical drop breakup events. Nevertheless, the following results provide preliminary insights into the dynamics of drop breakup above breaking surface waves.

**Interscale droplet-mass flux**

The time variation of the measured interscale droplet-mass flux is plotted in Figure 8. In the case of turbulent bubble breakup, the bubble breakup flux exhibited two characteristic time intervals: an initial time interval where the flux oscillated about a low mean value, and a subsequent interval where the flux increased with time (Chan *et al.*, 2021c). Here, the droplet breakup flux exhibits a clear peak shortly after touchdown of the overturning wave surface, but then eventually decays with time. This is suggestive of a lack of multigenerational drop breakup unlike in the corresponding bubble breakup process, which is corroborated by the smaller size range spanned by the drop population as noted earlier.

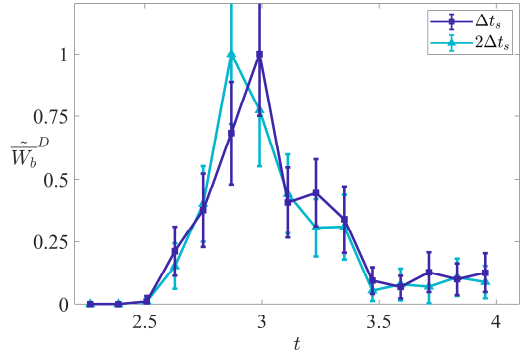

**Figure 8:** Time variation of the interscale droplet-mass flux, plotted for the two choices of the snapshot interval used in the drop tracking algorithm, volume averaged over the domain of each realization, ensemble averaged over all relevant realizations, and further averaged over the drop-size range $[9.74 \times 10^{-3}, 1.86 \times 10^{-2}]$. The flux is normalized by its maximum value in each data series. The uncertainties in the flux span two standard errors in each direction.

**Interscale locality of the droplet-mass flux**

The interscale locality of the droplet-mass flux is plotted for two cutoff sizes in Figures 9–12. These two cutoff sizes fall within the size range used to plot Figure 8 and approximately within the power-law scaling region observed in Figure 7.

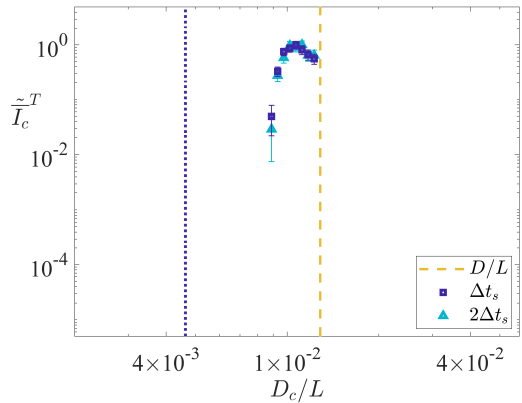

**Figure 9:** Ultraviolet locality of the interscale droplet-mass flux at the cutoff size $D/L = 1.28 \times 10^{-2}$, volume averaged over the domain of each realization, ensemble averaged over all relevant realizations, and time averaged over the interval $[2.30, 3.95]$. The vertical dotted line denotes the grid resolution, while the vertical dashed line denotes the cutoff size. The uncertainties in the distribution span two standard errors in each direction.

Figures 9 and 10 suggest that the degree of ultraviolet locality in the drop breakup process is relatively low, as a plateau is observed in the integrand near the cutoff size. While Figure 9 exhibits significant nonlocality since the integrand momentarily increases with distance from the cutoff size, the cutoff size is relatively close to the grid resolution, and Figure 10 is more representative of the ultraviolet locality properties of the drop breakup process.

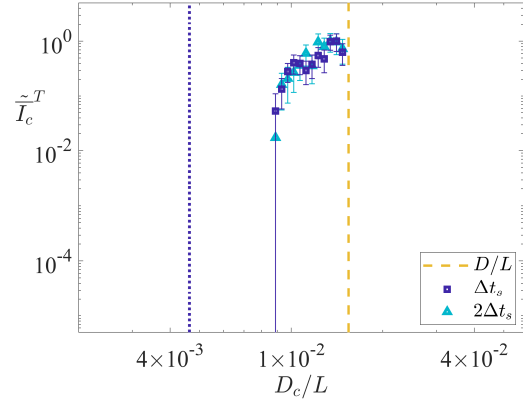

**Figure 10:** Ultraviolet locality of the interscale droplet-mass flux at the cutoff size $D/L = 1.54 \times 10^{-2}$, volume averaged over the domain of each realization, ensemble averaged over all relevant realizations, and time averaged over the interval $[2.30, 3.95]$. The vertical dotted line denotes the grid resolution, while the vertical dashed line denotes the cutoff size. The uncertainties in the distribution span two standard errors in each direction.

Physically, low ultraviolet locality suggests the generation of child drops of a broad range of sizes since the ultraviolet locality integrand extends to small child drop sizes with a low rate of decay. We revisit this observation in the discussion of Figures 14 and 15.

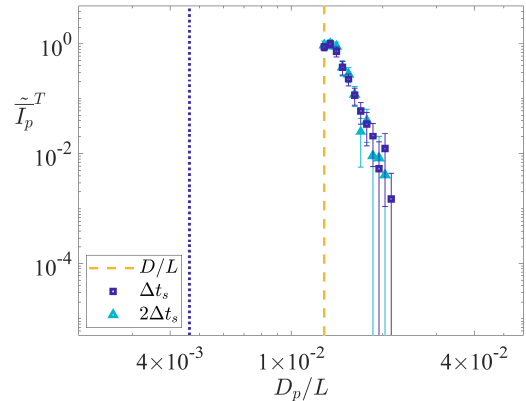

**Figure 11:** Infrared locality of the interscale droplet-mass flux at the cutoff size $D/L = 1.28 \times 10^{-2}$, volume averaged over the domain of each realization, ensemble averaged over all relevant realizations, and time averaged over the interval $[2.30, 3.95]$. The vertical dotted line denotes the grid resolution, while the vertical dashed line denotes the cutoff size. The uncertainties in the distribution span two standard errors in each direction.

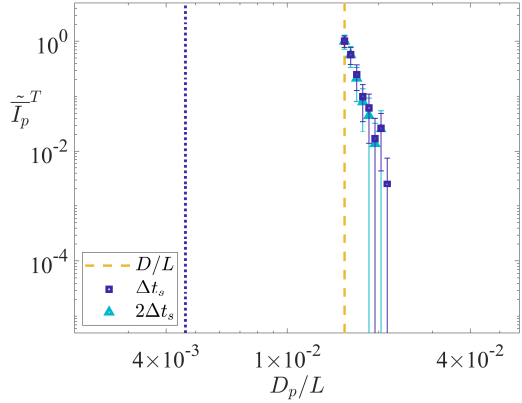

**Figure 12:** Infrared locality of the interscale droplet-mass flux at the cutoff size $D/L = 1.54 \times 10^{-2}$, volume averaged over the domain of each realization, ensemble averaged over all relevant realizations, and time averaged over the interval [2.30, 3.95]. The vertical dotted line denotes the grid resolution, while the vertical dashed line denotes the cutoff size. The uncertainties in the distribution span two standard errors in each direction.

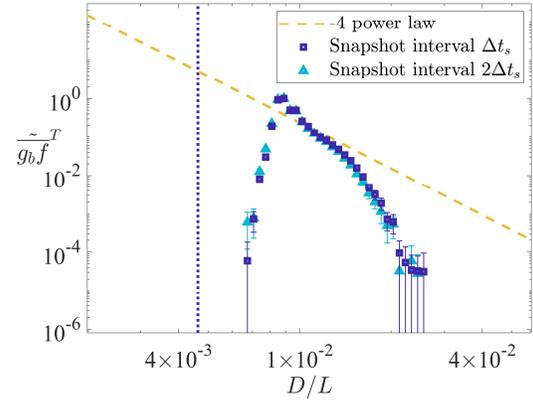

**Figure 13:** The normalized differential breakup rate, volume averaged over the domain of each realization, ensemble averaged over all relevant realizations, and time averaged over the interval [2.30, 3.95]. The vertical dotted line denotes the grid resolution, while the dashed line denotes the theoretical $D^{-4}$ power-law scaling associated with a size-invariant interscale flux. The uncertainties in the distribution span two standard errors in each direction.

Conversely, Figures 11 and 12 suggest that the degree of infrared locality in the drop breakup process is high. In both plots, the integrand decays quickly from the cutoff size. This is partly due to the corresponding rapid drop-off in the drop size distribution, which restricts the average size of the parent drop in breakup events. Significant infrared locality implies that the drop breakup process is unlikely to be sensitive to the large-scale geometric configuration of the wave. This supports an earlier remark that the statistics of sufficiently small drops are likely independent of $\text{Re}_L$ and $\text{We}_L$ provided these parameters are sufficiently large and the breakup dynamics are sufficiently scale-local. Specifically, since the observed dynamics are infrared local, the results of this work, derived from a 27-cm breaking surface wave, may be of relevance to those generated by full-scale ship motions due to the weak dependence of these statistics on the large-scale geometry of the problem.

**Drop breakup probabilities and rates**

We now turn to the constituent quantities of the locality integrands, $g_b$ and $q_b$. Note that $g_b$ cannot be directly measured in the simulation ensemble due to the nature of the averaging operators, and $g_b f$ is presented instead. Since the averaging operators act on the product $g_b f$, the corresponding probability distribution of child drop sizes is denoted $\check{q}_b$.

Figure 13 plots the differential breakup rate, whose shape resembles that of the drop size distribution in Figure 7, except for a short sloped plateau at intermediate sizes where a $D^{-4}$ power-law fit is appropriate. This power-law scaling is representative of a constant interscale flux. Recall the expression (3) for a local interscale flux. If we write $v_D \sim g_b D$, then it is apparent that $g_b f \sim D^{-4}$ corresponds to a size-invariant flux. The range of sizes where this is present in Figure 13 is limited, and ensembles of more resolved simulations are necessary to confirm it. A size-invariant interscale flux supports the claim that the drop breakup process is insensitive to the large-scale geometric configuration of the wave.

Figures 14 and 15 plot the probability distribution of child drop volumes, conditioned on two different sets of parent drop sizes. The distributions indicate that as the parent drop size increases, large-size-ratio breakup events are favored. Taken together with Figure 13, the probability distributions suggest that capillary effects are at play here. The rupture of liquid sheets and tubes into drops whose characteristic sizes are of the same order as the sheet thickness or tube diameter will generate large-size-ratio events for the initial breakup events, where the sheet/tube is mostly intact and has a larger equivalent diameter than the child drops. Smaller-size-ratio events are then generated for the final events where the sheet/tube has mostly ruptured and equal-volume binary breakup is more likely. Further investigation of the geometry of the parent drops of individual breakup events is necessary to confirm this hypothesis. The significant uncertainties in Figures 14 and 15 also necessitate larger simulation ensembles to improve statistical convergence.

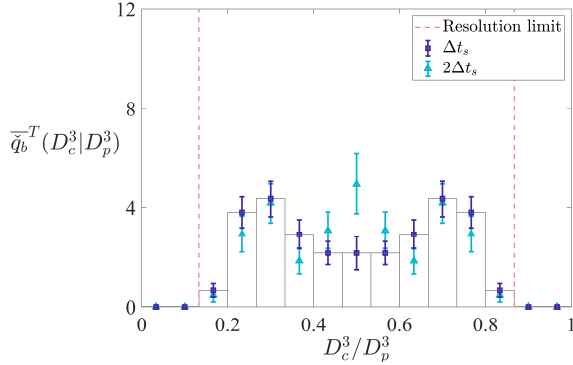

**Figure 14:** Probability distribution of child drop volumes, volume averaged over the domain of each realization, ensemble averaged over all relevant realizations, time averaged over the interval [2.30, 3.95], and averaged over parent drops of sizes [0.0128, 0.0154]. The vertical dashed lines denote child drops of sizes equal to the grid resolution. The uncertainties in the distributions span one standard error in each direction.

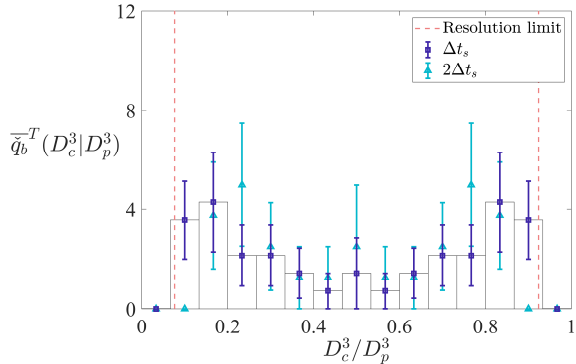

**Figure 15:** Probability distribution of child drop volumes, volume averaged over the domain of each realization, ensemble averaged over all relevant realizations, and time averaged over the interval [2.30, 3.95], and averaged over parent drops of sizes [0.0154, 0.0186]. The vertical dashed lines denote child drops of sizes equal to the grid resolution. The uncertainties in the distributions span one standard error in each direction.

## CONCLUSIONS

Drops of a broad range of sizes are generated by breaking surface waves with significant impact on our weather and climate, as well as ship operation and performance. Using an analysis toolkit comprising theoretical population balance analysis, a numerical breaking-wave simulation ensemble, and structure identification and tracking algorithms, the interscale nature of drop breakup in breaking waves is probed. The bubble and drop size distributions are used to validate the simulation ensemble, although additional investigations into the size distribution produced by the relatively new numerical method are necessary, and more resolved simulation ensembles are required to obtain more extended size ranges for the expected power-law scalings.

The interscale droplet-mass flux exhibits a single peak in time, suggesting the absence of multigenerational drop breakup unlike in the case of the turbulent bubble breakup cascade. The flux is demonstrated to be ultraviolet nonlocal but infrared local. This implies that the drop breakup process is unlikely to be sensitive to the large-scale geometric configuration of the wave, and general SGS models remain a viable possibility. The $D^{-4}$ power-law scaling at an intermediate range of sizes in the differential breakup rate supports this insensitivity to large-scale behavior, since it implies that the interscale breakup flux is size invariant in this size range.

Ultraviolet nonlocality of the interscale flux suggests that a breakup cascade mechanism is absent and geometric dependence may be present. The variation of the probability distribution of child drop volumes with parent drop size suggests that capillary breakup is at play. Capillary breakup of high-aspect-ratio structures, such as sheets and tubes, can initially generate large-size-ratio breakup events and eventually smaller-size-ratio breakup events as their volumes dwindle, resulting in both a parent-drop-size dependence in the child drop volume distribution and in ultraviolet nonlocality due to the broad range of size ratios present in the breakup events. This hypothesis needs to be confirmed with an investigation of the geometry of the parent drops, as well as more resolved simulation ensembles with a larger number of realizations for statistical convergence.

## ACKNOWLEDGEMENTS

The author is grateful to Prof. Parviz Moin, Prof. Perry Johnson, Dr. Javier Urzay, and Dr. Michael Dodd for previous guidance on this work. The foundations of this work were supported by the U.S. Office of Naval Research (ONR), as well as the Agency for Science, Technology and Research, Singapore, and Stanford University. ONR program managers included Dr. Ki-Han Kim and Dr. Thomas Fu. The author would also like to acknowledge computational resources from the U.S. Department of Energy's INCITE and ALCC Programs, as well as the Advanced Simulation and Computing program of the U.S. Department of

## APPENDIX: GRID SENSITIVITY

Here, we present the bubble and drop size distributions from the baseline ensemble from which the results above were derived, in comparison with a more resolved ensemble with twice the grid resolution containing 15 realizations, in order to examine grid sensitivity of the bubble and drop statistics.

### Bubble size distribution

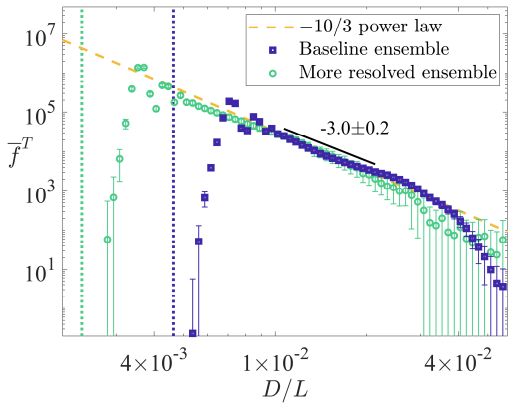

**Figure A1:** Bubble size distribution computed from two breaking-wave simulation ensembles, volume averaged over the domain of each realization, ensemble averaged over all relevant realizations, and time averaged over the interval [2.30, 3.14]. The vertical dotted lines denote the grid resolutions of the two ensembles, while the dashed line denotes the analytically derived and previously observed $D^{-10/3}$ power-law scaling. The uncertainties in the distribution and scaling exponent span two standard errors in each direction.

As observed in previous work (Chan *et al.*, 2021c), the bubble size distribution is not sensitive to the grid resolution at the considered grid size. Note that Figure A1 is obtained using an algebraic volume-of-fluid method, in contrast to the size distributions of Chan *et al.* (2021c), which were obtained using a geometric volume-of-fluid method. The former has a larger effective interface thickness and nominally requires higher resolution for comparable accuracy. Nevertheless, at the considered grid resolution, grid insensitivity is observed in this work even for the former method, thus providing a benchmark measurement for the algebraic volume-of-fluid method applied to multiscale multiphase flows.

### Drop size distribution

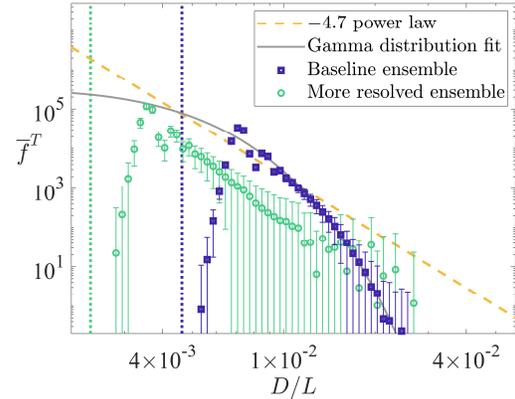

**Figure A2:** Drop size distribution computed from two breaking-wave simulation ensembles, volume averaged over the domain of each realization, ensemble averaged over all relevant realizations, and time averaged over the interval [2.30, 3.95]. The vertical dotted lines denote the grid resolutions of the two ensembles, while the dashed line denotes a $D^{-4.7}$ power-law scaling close to that observed in previous works. The uncertainties in the distribution span two standard errors in each direction.

Conversely, the drop size distribution is observed to be sensitive to the grid resolution at the considered grid size, although the distributions from both ensembles still compare well with the $D^{-4.7}$ power-law scaling observed in previous works. The larger number of intermediate-sized drops in the baseline ensemble is likely attributable to the persistence of liquid ligaments with blobs, which have a higher tendency to be broken up in the more resolved ensemble. The numerical connection of these blobs can have a significant effect on the computation of the resulting size distribution (Chan *et al.*, 2021a).

The persistence of connected liquid ligaments may be explained as follows: with a lower grid resolution, thinner features cannot be resolved as effectively, including the drainage film linking two liquid blobs that are about to break. A thinner film is associated with a larger drainage velocity and thus a greater tendency to rupture. While the breakup of bubbles also requires drainage of the liquid film separating two bubbles, the same issue is not observed for bubble breakup since the shapes of sufficiently large air masses are largely governed by deformation from the motion of the surrounding inertial liquid and capillary breakup plays a secondary role in terms of the time to breakup. This is corroborated by the observation that the inverse eddy turnover time is an accurate predictor of the bubble breakup frequency in a turbulent bubble breakup cascade (Chan *et al.*, 2021bc). On the other hand, the inertial fluid is internal to the liquid masses in the case of drop breakup. While a turbulent breakup cascade may still exist, the characteristic large scale of the cascade would then be governed by the size of the encompassing liquid mass rather than of the wave itself. Under such a constraint, viscous drainage and capillary breakup would play a stronger role in defining the structure breakup time, and numerical resolution of the connecting film would be crucial to resolving the breakup process. Such a physical argument corroborates the observation of ultraviolet nonlocal drop breakup described earlier in this work.

**REFERENCES**

Andreas, E.L., "Sea spray and the turbulent air-sea heat fluxes," Journal of Geophysical Research: Oceans, Vol. 97, 1992, pp. 11429–11441.

Andreas, E.L., Edson, J.B., Monahan, E.C., Rouault, M.P. and Smith, S.D., "The spray contribution to net evaporation from the sea: a review of recent progress," Boundary-Layer Meteorology, Vol. 72, 1995, pp. 3–52.

Anguelova, M., Barber Jr., R.P. and Wu, J., "Spume drops produced by the wind tearing of wave crests," Journal of Physical Oceanography, Vol. 29, 1999, pp. 1156–1165.

Andreas, E.L. and Emanuel, K.A., "Effects of sea spray on tropical cyclone intensity," Journal of the Atmospheric Sciences, Vol. 58, 2001, pp. 3741–3751.

Chan, W.H.R., Urzay, J. and Moin, P., "Subgrid-scale modeling for microbubble generation amid colliding water surfaces," Proceedings of the 32$^{nd}$ Symposium on Naval Hydrodynamics, 2018, Hamburg, Germany.

Chan, W.H.R., Mirjalili, S., Jain, S.S., Urzay, J., Mani, A. and Moin, P., "Birth of microbubbles in turbulent breaking waves," Physical Review Fluids, Vol. 4, 2019, 100508.

Chan, W.H.R., "The bubble breakup cascade in turbulent breaking waves and its implications on subgrid-scale modeling," Ph.D. dissertation, Stanford University, 2020.

Chan, W.H.R., Dodd, M.S., Johnson, P.L. and Moin, P., "Identifying and tracking bubbles and drops in simulations: a toolbox for obtaining sizes, lineages, and breakup and coalescence statistics," Journal of Computational Physics, Vol. 432, 2021a, 110156.

Chan, W.H.R., Johnson, P.L. and Moin, P., "The turbulent bubble break-up cascade. Part 1. Theoretical developments," Journal of Fluid Mechanics, Vol. 912, 2021b, A42.

Chan, W.H.R., Johnson, P.L., Moin, P. and Urzay, J., "The turbulent bubble break-up cascade. Part 2. Numerical simulations of breaking waves," Journal of Fluid Mechanics, Vol. 912, 2021c, A43.

Chang, S., "Ships cause their own stormy seas," Physics Today, Vol. 70, 2017, pp. 20–21.

Cunliffe, M., Engel, A., Frka, S., Gašparović, B., Guitart, C., Murrell, J.C., Salter, M., Stolle, C., Upstill-Goddard, R. and Wurl, O., "Sea surface microlayers: A unified physicochemical and biological perspective of the air–ocean interface," Progress in Oceanography, Vol. 109, 2013, pp. 104–116.

de Leeuw, G., "Vertical profiles of giant particles close above the sea surface," Tellus, Vol. 38B, 1986, p. 51–61.

de Leeuw, G., Neele, F.P., Hill, M., Smith, M.H. and Vignati, E., "Production of sea spray aerosol in the surf zone," Journal of Geophysical Research: Atmospheres, Vol. 105, 2000, pp. 29397–29049.

de Leeuw, G., Andreas, E.L., Anguelova, M.D., Fairall, C.W., Lewis, E.R., O'Dowd, C., Schulz, M. and Schwartz, S.E., "Production flux of sea spray aerosol," Reviews of Geophysics, Vol. 49, 2011, RG2001.

Deike, L., "Mass transfer at the ocean–atmosphere interface: the role of wave breaking, droplets, and bubbles," Annual Review of Fluid Mechanics, Vol. 54, 2022, pp. 191–224.

Erinin, M.A., Wang, S.D., Liu, R., Towle, D., Liu, X. and Duncan, J.H., "Spray generation by a plunging breaker," Geophysical Research Letters, Vol. 46, 2019, pp. 8244–8251.

Erinin, M.A., Néel, B., Ruth, D.J., Mazzatenta, M., Jaquette, R.D., Veron, F. and Deike, L., "Speed and acceleration of droplets generated by breaking wind-forced waves," Geophysical Research Letters, Vol. 49, 2022, e2022GL098426.

Fairall, C.W., Banner, M.L., Peirson, W.L., Asher, W. and Morison, R.P., "Investigation of the physical scaling of sea spray spume droplet production," Journal of Geophysical Research: Oceans, Vol. 114, 2009, C10001.

Fan, J., Wang, Y., Rosenfeld, D. and Liu, X., "Review of aerosol–cloud interactions: mechanisms, significance, and challenges," Journal of the Atmospheric Sciences, Vol. 73, 2016, pp. 4221–4252.

Kim, D., Ivey, C.B., Ham, F.E. and Bravo, L.G., "An efficient high-resolution Volume-of-Fluid method with low numerical diffusion on unstructured grids," Journal of Computational Physics, Vol. 446, 2021, 110606.

Koga, M., "Direct production of droplets from breaking wind-waves—its observation by a multi-colored overlapping exposure photographing technique," Tellus, Vol. 33, 1981, pp. 552–563.

Mehta, S., Ortiz-Suslow, D.G., Smith, A.W. and Haus, B.K., "A laboratory investigation of spume generation in high winds for fresh and seawater," Journal of Geophysical Research: Atmospheres, Vol. 24, 2019, pp. 11297–11312.

Mostert, W., Popinet, S. and Deike, L., "High-resolution direct simulation of deep water breaking waves: transition to turbulence, bubbles and droplet production," Journal of Fluid Mechanics, Vol. 942, 2022, A27.

Smith, M.H., Park, P.M. and Consterdine, I.E., "Marine aerosol concentrations and estimated fluxes over the sea," Quarterly Journal of the Royal Meteorological Society, Vol. 119, 1993, pp. 809–824.

Tao, W.-K., Chen, J.-P., Li, Z., Wang, C. and Zhang, C., "Impact of aerosols on convective clouds and precipitation," Reviews of Geophysics, Vol. 50, 2012, RG2001.

Veron, F., Hopkins, C., Harrison, E.L. and Mueller, J.A., "Sea spray spume droplet production in high wind speeds," Geophysical Research Letters, Vol. 39, 2012, L16602.

Veron, F., "Ocean Spray," Annual Review of Fluid Mechanics, Vol. 47, 2015, pp. 507–538.

Villermaux, E., "Fragmentation," Annual Review of Fluid Mechanics, Vol. 39, 2007, pp. 419–446.

Wang, Z., Yang, J. and Stern, F., "High-fidelity simulations of bubble, droplet and spray formation in breaking waves," Journal of Fluid Mechanics, Vol. 792, 2016, pp. 307–327.

Wu, J., "Spray in the atmospheric surface layer: review and analysis of laboratory and oceanic results," Journal of Geophysical Research: Oceans, Vol. 84, 1979, pp. 1693–1704.

Wu, J., Murray, J.J. and Lai, R.J., "Production and distributions of sea spray," Journal of Geophysical Research: Oceans, Vol. 89, 1984, pp. 8163–8169.

## DISCUSSION

Martin A. Erinin, Department of Mechanical and Aerospace Engineering, Princeton University

The authors are to be commended on their interesting paper on the study of droplet generation in breaking water waves. The study of droplet generation mechanisms in breaking waves is a transient and multi-phase phenomena that is challenging to study experimentally and numerically. In this paper, the author uses a recently developed theoretical framework to study multiphase phenomena and applies it to study the droplet generation problem. Analysis of this framework is carried out on a large set of numerical simulations of breaking waves, which were conducted by the author.

I noticed that the author mentions that there are significantly fewer droplets detected in the simulations compared to the number of bubbles. Can the author comment on this observation? Does the author think this is because of simulation resolution limits or otherwise? As the author notes, the average droplet size is typically smaller than the average bubble size. Does this mean that for the bubble population resolved by the current simulations, there is potentially a large number of droplets that would have been produced by these bubbles if the resolution were increased? Can the author comment on how this may impact the analysis?

Additionally, breaking waves are known to produce droplets by splashing and bubble popping. However, the relative contribution of the two mechanisms is not well understood or characterized. Can the author comment on the viability of distinguishing between splashing and bubble popping events in these simulations? How would these potentially different spray generation mechanisms impact the theoretical framework presented by the author?

## AUTHOR'S REPLY

Thank you for your comments and questions, each of which is addressed below.

1. There are limitations due to resolution limits as discussed in the appendix, but higher resolution simulations do not yield significantly more drops as the moment. It may be expected, however, that further resolution will eventually result in a larger drop population once the large-drop population is grid converged, in accordance with corresponding trends in the small- and large-bubble populations.

   The observed size difference is in line with experimental measurements that also reveal a physical discrepancy between the average droplet and bubble sizes. More resolved simulations will be pursued as they become feasible to obtain more converged statistics.

2. The statistics themselves will likely not reveal differences between these two types of events at face value. The current algorithm is also limited to binary breakup events (although it can be extended to polyadic events) and this presents limitations in directly discerning between the two types.

   However, the two types of events can be distinguished by additional constraints, such as whether a coalescence event happened before the breakup event in its vicinity (in the case of splashing). This has not been carried out in the current work and will be considered in future follow-up investigations.

## DISCUSSION

Yinghe Qi, Department of Mechanical Engineering, Johns Hopkins University

This work focuses on the interscale physics of droplet breakup in breaking wave by using a framework including numerical simulation, population balance analysis and tracking algorithm. The definition of the interscale mass flux enable the author to quantify the contribution to the local mass transfer from multiple scales. It is shown that the mass flux is ultraviolet nonlocal and infrared local, suggesting the negligible contribution from large scale wave configuration to the local breakup process. The author also proposes different mechanisms for drop breakup which will then affect the distribution of the child bubble.

As a discusser, I was impressed not only by the significance of this work, but also by the insightful analysis regarding the locality of mass transfer. I would like to recommend the publication of this paper after the following comments are addressed.

The main comment: has the author thought about the relationship between ultraviolet and infrared locality? For example, if we assume the scaling for ultraviolet case is alpha (suggested by figure 10, a bit further from the grid resolution) and the scaling for the infrared case is beta (by figure 11 and 12, the scaling seems very similar), then I wouldn't be suppressed if alpha and beta are connected somehow. This relationship might be found through the population balance equation together with the scaling of drop number density (-4.7 in this paper). But I do understand that this relationship might be complicated given the PDF of the child drop changes as a function of size (figure 14 and 15). It would also be helpful if the author could include some comparison between this work and the mass flux for bubble cascade (from another paper by the same author).

Some minor comments:

In equation 3, the author is encouraged to elaborate more on the characteristics speed to help people understand this term. Intuitively this speed should be proportional to the breakup frequency, which is also an assumption in the discussion of figure 13.

Figure 8 does not seem to sufficiently support the argument of the lack of mutigenerational breakup. The horizontal axis can be normalized by some averaged breakup frequency to illustrate this point.

Some snapshots can be provided to illustrate the breakup mechanism (for large and small size ratio).

**AUTHOR'S REPLY**

Thank you for your comments and questions, each of which is addressed below.

1. For a self-similar flux, if $I_p \sim D_p^\beta$, $I_c \sim D_c^\alpha$, and $W_b \sim D^0$, then $D^{\alpha+1} D^{\beta+1} \sim D^0$, or $\alpha + \beta = -2$. The identity is generally satisfied in the case of the turbulent bubble breakup cascade (Chan *et al.*, 2021c). However, it is challenging to discern here because (a) the power-law trends in Figures 9 and 10 are not as clear due to the lower degree of ultraviolet locality and (b) the self-similar size range in Figure 13 is limited. More resolved simulations will be necessary to further probe this relation.

2. As the reviewer has remarked, the characteristic speed is portrayed as proportional to the breakup frequency as noted in the discussion of Figure 13. This is less of an assumption but more of a consequence of scaling analysis. A remark to clarify the relation between $v_D$ and $g_b$ has been added to the discussion of Equation 3.

3. The horizontal axis has already been normalized by $\sqrt{L/g}$, which is representative of a large-eddy turnover time. The single peak and the width of the peak suggest that the breakup dynamics have largely ceased after about one large-eddy turnover time. It is true that this does not definitively imply the lack of multigenerational breakup since the average lifetime of a cascade is also about one large-eddy turnover time, but Figure 8 should be contrasted with the corresponding behavior of bubbles, where the breakup behavior is more prolonged with a two-stage process (Chan *et al.*, 2021c).

4. The focus of this work is on event statistics rather than individual events themselves, but analysis of the breakup mechanism for individual events will be embarked upon as more resolved simulations become more feasible with computational advancements and small-drop dynamics become more reliable with additional resolution.